\documentclass[twocolumn]{aastex631}
\usepackage{{booktabs}}
\usepackage{romannum}
\usepackage{float}
\usepackage{hyperref}
\usepackage{soul}
\usepackage{xfrac}
\usepackage{acronym}
\setulcolor{red}
\setstcolor{red}

\def\rsun{R$_{\odot}$}

\newcommand{\ca}{\mbox{Ca\,{\sc ii}~K\,}}
\newcommand{\cah}{\mbox{Ca\,{\sc ii}~H\,}}
\newcommand{\pnikso}{PNI$_{\rm KoSO}$}
\newcommand{\pnipspt}{PNI$_{\rm PSPT-R}$}
\newcommand{\Bp}{$B_{\rm P}$}
\newcommand{\Bwso}{$B_{\rm P}^{\rm WSO}$}
\newcommand{\Baft}{$B_{\rm P}^{\rm AFT}$}

\graphicspath{ {./figures/} }
\renewcommand{\textbf}[1]{\textcolor{magenta}{{\bf #1}}}

\begin{document}
\acrodef{sse}[SSE]{standard statistical error}
\def\sse{\ac{sse}}

\acrodef{cc}[CC]{correlation coefficient}
\def\cc{\ac{cc}}

\acrodef{mdi}[MDI]{Michelson Doppler Imager}
\def\mdi{\ac{mdi}}

\title{\ca\ Polar Network Index of the Sun: A Proxy for Historical Polar Magnetic Field}


\shorttitle{The polar field of the Sun}
\shortauthors{D. K. Mishra et al.}

\correspondingauthor{Bibhuti Kumar Jha}
\email{maitraibibhu@gmail.com}

\author[0009-0003-1377-0653]{Dibya Kirti Mishra}
\affiliation{Aryabhatta Research Institute of Observational Sciences, Nainital-263002, Uttarakhand, India}
\affiliation{Mahatma Jyotiba Phule Rohilkhand University, Bareilly-243006, Uttar Pradesh, India}

\author[0000-0003-3191-4625]{Bibhuti Kumar Jha}
\affiliation{Southwest Research Institute, Boulder, CO 80302, USA}

\author[0000-0002-0335-9831]{Theodosios Chatzistergos}
\affiliation{Max Planck Institute for Solar System Research, Justus-von-Liebig-Weg 3, D-37077 Göttingen,Germany}

\author[0000-0003-2596-9523]{Ilaria Ermolli}
\affiliation{INAF Osservatorio Astronomico di Roma, Via Frascati 33, 00078 Monte Porzio Catone, Italy}

\author[0000-0003-4653-6823]{Dipankar Banerjee}
\affiliation{Aryabhatta Research Institute of Observational Sciences, Nainital-263002, Uttarakhand, India}
\affiliation{Indian Institute of Astrophysics, Koramangala, Bangalore 560034, India}
\affiliation{Center of Excellence in Space Sciences India, IISER Kolkata, Mohanpur 741246, West Bengal, India}

\author[0000-0003-0621-4803]{Lisa A. Upton}
\affiliation{Southwest Research Institute, Boulder, CO 80302, USA}

\author{M. Saleem Khan}
\affiliation{Mahatma Jyotiba Phule Rohilkhand University, Bareilly-243006, Uttar Pradesh, India}

\begin{abstract}

The Sun's polar magnetic field is pivotal in understanding solar dynamo processes and forecasting future solar cycles. However, direct measurements of the polar field is only available since the 1970s. The chromospheric \ca\ polar network index (PNI; the fractional area of the chromospheric network regions above a certain latitude) has recently emerged as a reliable proxy for polar magnetic fields. In this study, we derive PNI estimates from newly calibrated, rotation-corrected \ca\ observations from the Kodaikanal Solar Observatory (1904–2007) and modern data from the Rome Precision Solar Photometric Telescope (2000–2022). We use both of those \ca archives to identify polar network regions with an automatic adaptive threshold segmentation technique and calculate the PNI. The PNI obtained from both the archives shows a significant correlation with the measured polar field from WSO (Pearson correlation coefficient $r>0.93$) and the derived polar field based on an Advective Flux Transport Model ($r>0.91$). The PNI series also shows a significant correlation with faculae counts derived from Mount Wilson Observatory observations ($r>0.87$) for both KoSO and Rome-PSPT data. Finally, we use the PNI series from both archives to reconstruct the polar magnetic field over a 119-year-long period, which includes last 11 solar cycles (Cycle~14\,--\,24). We also obtain a relationship between the amplitude of solar cycles (in 13-month smoothed sunspot number) and the strength of the reconstructed polar field at the preceding solar cycle minimum to validate the prediction of the ongoing solar cycle, Cycle\,25.
\end{abstract}


\section{Introduction} \label{sec:intro}

The polar magnetic field serves as a critical component within the dynamo models, subsequently giving rise to the toroidal field and the manifestation of sunspots \citep[see, e.g.][]{charbonneau_dynamo_2020}. This polar field exhibits an antiphase relationship with the solar cycle in which it attains maximal values near the sunspot cycle minima. Notably, this field holds pivotal importance in forecasting the amplitude of the forthcoming solar cycles \citep{schatten1978}, with a remarkable correlation observed between sunspot maximum and polar field strength at the minimum preceding the solar cycle \citep{jiang2007,wang2009,kitchatinov2011,munoz2013,petrovay2020,pawan2021,pawan2022,upton2023JGRA}. Recent efforts, as highlighted by \citet{bhowmik2018,janssens2021,jha2024}, have utilized polar field information for predicting the amplitude of the ongoing Solar Cycle\,25 and the timing of the polar field reversal. 
During solar minimum, the polar magnetic field of each hemisphere exhibits a predominantly unipolar configuration. 
This is typically measured as the average field strength in a specified latitude range.
The latitudinal observation range, however, differs among different studies.
For example, the Wilcox Solar Observatory (WSO) provides a time series of polar field that corresponds to approximately \citep[$\pm(55\degr-90\degr)$; ][]{svalgaard_strength_1978} band,  \citet{sun_polar_2015} used $\pm(60\degr-90\degr)$, whereas \citet{Upton2014} used three bands in their study i.e. $\pm(55\degr-90\degr)$, $\pm(70\degr-90\degr)$ and $\pm(85\degr-90\degr)$. 
This choice of latitude range is again different for the reproduction of the polar field from polar faculae, which is typically considered within $\pm(70\degr-90\degr)$ \citep[e.g.][]{Munoz2012,hovis-afflerbach_two_2022}. The average strength of the polar field measured by the Helioseismic and Magnetic Imager \citep[HMI;][]{scherrer2012} line of sight magnetograms \citep{janardhan2018A&A} was $\approx$5\,G peaking near solar minimum following Cycle 24  (year 2017); however, small-scale facular features with strengths in the kilogauss (kG) range have also been observed in the polar regions \citep{petrie2015LRSP}.

Measuring the polar magnetic field of the Sun from the Sun-Earth line presents significant challenges due to the foreshortening and relatively weaker polar field. Despite that, the initial attempt to measure the polar magnetic field was pioneered by \citet{babcock1955} using solar magnetographs, but the systematic measurements of the polar field only started in 1976 by WSO \citep{svalgaard1978,hoeksema_structure_1984,sanderson_observations_2003}.
Although magnetograms over earlier periods exist, e.g. from Kitt Peak \citep[][covering 1974--2003]{livingston_kitt_1976} and Mount Wilson Observatory \citep[MWO;][covering 1967--2013]{ulrich_mount_2002}, their instruments underwent multiple changes, making their data not optimal for studying the long term evolution of the polar magnetic field. 

The lack of polar field measurements before 1976 necessitates the search for other proxies for the polar fields, such as photospheric polar faculae \citep[hereafter, polar faculae;][]{makrov2001,petrie2015,hovis-afflerbach_two_2022,elek_exploring_2024}, which extend the record of the polar field estimates back to the year 1906. 
In addition to polar faculae, polar filaments have been investigated as a polar field proxy in some recent studies \citep{diercke_chromospheric_2019,xu_migration_2021,chatzistergos_analysis_2023}.
 Among all the proxies,  polar faculae are the most widely used for estimating polar field strength due to their continuous observations and the consistent availability of polar faculae count. For instance, polar faculae counts have been used to estimate the polar fields employing historical white-light observation from the MWO \citep[1906\,--\,2007;][]{sheeley1991,sheeley_century_2008,Munoz2012,munoz2013}. However, all existing records of polar faculae suffer from potential bias due to the manual identification of polar faculae, which is most prominent when counted across observatories. 
For example, \citet{li2002} has found that the polar faculae count varies between MWO and the National Astronomical Observatory of Japan (NAOJ)  for Cycle\,20 and Cycle\,21. 


Potentially, a better proxy than polar faculae for the polar fields can be derived from solar chromospheric observations in the \ca\ line. 
That is because of the strong relationship between \ca\ brightness and solar surface magnetic field strength \citep[see][and references therein]{chatzistergos_recovering_2019,chatzistergos_full-disc_2022}. Furthermore, chromospheric network regions can easily be detected in \ca\ observations, also making the network appear in polar regions  (hereafter, polar network). Recent work with \cah\ data from the Solar Optical Telescope (SOT) onboard the Hinode spacecraft, having a high spatial resolution of 0.32\arcsec to 0.65\arcsec, has also highlighted the potential of polar network bright points as a valuable indicator for estimating polar magnetic fields \citep{narang2019}. These polar bright points have been observed with magnetic fields exceeding 1\,kG \citep{tritschler2002}, and they have been noted to correlate well with photospheric bright faculae \citep{kaithakkal2013,narang2019}. In an earlier study, \citet{makrov1989} reported a linear relationship between polar faculae and \ca\ bright points (for the period of 1940\,--\,1957), along with sunspot areas in each hemisphere, from analysis of polar faculae data from the Pulkovo Observatory and \ca\ bright points in Kodaikanal Solar Observatory (KoSO) observations.

More recently, \citet{priyal2014} suggested the use of polar network index (PNI), which is calculated as the number of \ca\ polar bright network pixels counted in the latitude range of $70^{\circ}-90^{\circ}$. \citet{priyal2014} analyzed \ca\ data from the KoSO \citep[1909\,--\,1990;][]{priyal2014SoPh}, and used a fixed threshold to identify the polar bright network regions in the mentioned latitude range. By constructing the PNI series for KoSO data, they established a strong correlation coefficient (0.95) between the PNI and the polar field as measured by WSO in the overlapping period of 1976 to 1990. 
However, later on, it was discovered that certain portions of the KoSO \ca\ data analyzed in that study suffer from various inconsistencies, which may significantly impact the estimation of the polar field. For example, \citet{Jha2022thesis, Jha2024prep} have discovered that a significant fraction of KoSO \ca\ observation has incorrect timestamp because of various reasons, which has led to the incorrect orientation of the image in previous studies. Considering that the accuracy of the pole definition relies entirely on the correct orientation of the solar disk in the image, such errors in orienting the images could have a significant impact on the PNI calculation. Moreover, the recent implementation of calibration techniques by \citet{theo2018,theo2019,theo2019SoPh,chatzistergos_analysis_2020} has significantly improved the quality of calibrated data and has opened the possibilities to detect polar network bright points more accurately.  





In this article, we use the newly calibrated and accurately rotation-corrected KoSO \ca\ observations from 1904 to 2007 together with modern Rome-PSPT (Rome Precision Solar Photometric Telescope) \ca observations from 2000\,--\,2022 to derive an accurate composite \ca\ PNI series covering 1904--2022. 
Based on this, we also recovered the polar magnetic field over the same period, covering 11 solar cycles (Cycle~14\,--\,24). 
The data generated by this study will be valuable assets for the long-term study of the Sun, as it will help constrain the reconstruction of polar fields generated by solar dynamo or surface flux transport (SFT) models.  In Section \ref{Sec:data}, we discuss the data utilized in this work; in Section \ref{Sec:method}, we outline our methodology for the detection of \ca\ network and the estimation of the polar magnetic field; subsequently, in Section \ref{Sec:results}, we present our findings and comparison of our estimated polar field with the existing polar field measurement. Finally, we summarize our conclusions in Section \ref{Sec:summary}. 


\section{Data} \label{Sec:data}

\begin{figure*}[htb!]
 \centering
\includegraphics[width=15cm]{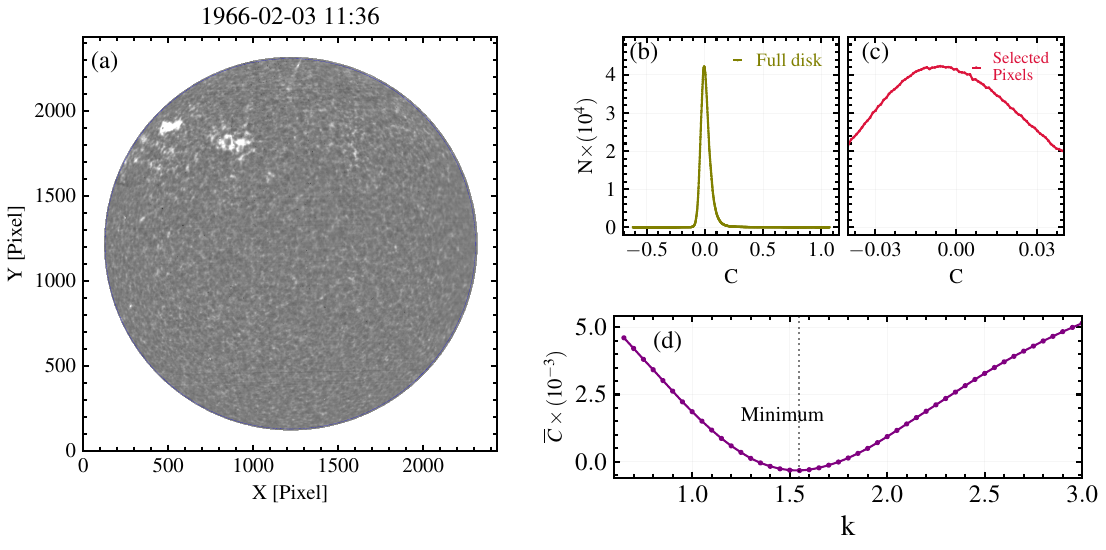}
\caption{(a) An illustrative example of the rotation-corrected \ca\ image (contrast $\in [-0.5,0.5]$) of KoSO observed on 1966-02-03 11:36 IST (06:06 UT). The blue circle shows the boundary at 0.98 \rsun\, which signifies the extent of the selected solar disk for the analysis. (b) The intensity distribution histogram (C) across the entire solar disk (olive green) for the image showcased in (a).
The skewness towards the right-hand side indicates the presence of bright features such as plages and network regions on the solar disk. (c) Presents an exemplar histogram distribution of selected pixels (red curve) falling within the range ($\overline{C} \pm k*\sigma$) for $k=1.55$. 
(d) Exhibits the variation of mean contrast ($\overline{C}$) corresponding to different values of $k$, enabling the selection of the minimum value of $\overline{C}$ ($\overline{C}_{\mathrm{min}}$) for a specific $k$ value, alongside the corresponding minimum standard deviation ($\sigma_{\mathrm{min}}$).} 
\label{network_detection}
\end{figure*}

The KoSO possesses an extensive archive of \ca\ spectroheliograms, having been acquired with a nominal bandpass of 0.05\,nm centered at 393.367\,nm, with records dating back to 1904 and spanning until 2007 \citep{priyal2014SoPh, Chatterjee2016, chatzistergos_analysis_2020,Jha2022thesis}. Originally captured on photographic plates/films, these data have undergone digitization using a 4096\,$\times$\,4096 pixels CCD sensor with a 16-bit depth and resulting images with a pixel scale of $\approx$\,0.9\arcsec. This digitized dataset is now accessible to the scientific community via the KoSO data repository.\footnote{The digitized data can be accessed through \url{https://kso.iiap.res.in/data.}} Various calibration techniques have been implemented on these \ca\ observations \citep[e.g.;][]{priyal2014,Chatterjee2016}; however, significant advancements have been made through the work of \citet{theo2019SoPh}, resulting in an enhanced series of \ca\ data. Additionally, \citet{Jha2022thesis,Jha2024prep} developed a precise method for orienting KoSO images, addressing inaccuracies in timestamps for specific periods. For our present study, we utilize recently calibrated \citep{chatzistergos_analysis_2020} and correctly oriented KoSO observations \citep{Jha2022thesis,Jha2024prep} spanning the period 1904\,--\,2007.
A representative example of calibrated and correctly oriented KoSO observation 
is shown in \autoref{network_detection}(a).

Complementing the KoSO dataset, we incorporate \ca\ observations from the Rome-PSPT (PSPT-R hereafter)\footnote{Available at \url{https://www.oa-roma.inaf.it/pspt-daily-images-archive/}.} which uses a CCD camera and an interference filter with a bandwidth of 0.25\,nm centered at 393.37\,nm and span from 1996 to the present \citep{ermolli10.3389}. We used PSPT-R data for the period of 2000\,--\,2022 having a pixel scale of 2.0\arcsec /pixel \citep{chatzistergos_analysis_2020} to extend our analysis during periods when KoSO \ca\ data are unavailable. The PSPT-R images we use here were processed the same way as KoSO data \citep{theo2018,theo2019,chatzistergos_analysis_2020}.

In addition to \ca\ data, we also utilized white-light polar faculae counts from MWO \citep[1907\,--\,2007;][]{munoz2013} for the purpose of comparison. 

On the other hand, for the direct polar field measurement, we use WSO polar field data\footnote{Available at \url{http://wso.stanford.edu/Polar.html}.}, which started taking polar field measurements from 1976 to the present \citep{svalgaard1978}. The WSO measurements have been the most long-lived and consistent estimates of the polar field strength; however, they are based on a single-pixel reading, which samples different latitudes over the course of a year. Therefore, the visibility of the poles varies throughout the year due to changes in the tilt of the solar rotation axis toward the Earth/observer ($B_{0}$)
The $B_{0}$ angle, which varies between $\pm7.25\degr$ throughout the year, makes homogeneous observation of the solar poles from the Sun-Earth line unfeasible. 
We have also used the polar field derived from the Advective Flux Transport Model \citep[AFT;][]{Upton2014, Upton2014a,jha2024} along with WSO measurement.  
AFT assimilates line-of-sight magnetograms from the Michelson Doppler Imager \citep[MDI;][]{scherrer1995} and HMI instruments to produce the full 360\degr\ map (synchronic map) of the Sun. 
These maps are used to derive the polar field above 55\degr\, similar to WSO, for the period of 1996 on-wards. 
For details about the AFT, data assimilation and calculation of the polar field, see the original papers mentioned above.  It is important to note that the derivation of the polar field using the AFT is calibrated to the HMI observations. This results in measurements that are approximately 1.8 times greater in amplitude than those obtained using the WSO. The increased resolution also results in less latitudinal variability over the year. These combined effects result in a significant difference in the polar field strengths derived from WSO and AFT.

Since the optimal viewing of the northern and southern poles occurs only during August--September ($B_{0}\approx+7.25\degr$) and February--March ($B_{0}\approx-7.25\degr$), respectively \citep[see, e.g.,][]{petrie2015LRSP,janssens2021}, in this work, we have considered only these months when calculating the yearly average facular counts, as well as the polar field from WSO and AFT. 
To be consistent with the \ca data, we only use the KoSO and PSPT-R observations from these four months, with total images of 16,009 and 1980, respectively.



\section{Methodology}\label{Sec:method}



To identify the bright network regions, we employ the adaptive threshold technique described by \citet[][hereafter NR method]{nesme1996}, which was initially utilized for the detection of the quiet Sun regions and faculae in Meudon spectroheliograms. 
We chose the NR method based on the results by \citet{ermolli2007A&A,ermolli2009,phdthesis,theo2019}, who showed that this method performs better than others from the literature without introducing activity-dependent inconsistencies.
In particular, this method is preferred over constant threshold methods where a fixed threshold is applied uniformly over the solar disk \citep{singh2012,priyal2014}; adaptive threshold methods where the threshold is dynamically adjusted based on the mean contrast ($\overline{C}$) and standard deviation ($\sigma$) \citep{Chatterjee2016}; and multiple level tracking, where multiple thresholds are sequentially applied until bright magnetic features are identified \citep{bovelet2001}. 
It was found, in fact, that the adaptive threshold technique, such as \citet{Chatterjee2016}, is susceptible to variations induced by solar activity, adversely impacting the detection of network regions, whereas the constant threshold method, such as \citet{priyal2014} is not ideal for ground-based archival datasets due to inherent non-uniformity in data quality over the long observation period.

\subsection{Detection of Polar Network}
\label{sec:polarnetworkdetection}
The first step in identifying the polar network involves determining the quiet Sun intensity utilizing the NR method. Here, we restrict our background estimation and later network identification within 0.98\,$R$ of the solar disk (as represented by the blue line in \autoref{network_detection}(a) to limit the uncertainties associated with detecting the polar network near the limb. In \autoref{network_detection}(b), the distribution of contrast values (C) on the solar disk (within 0.98\,$R$; olive green) is illustrated. 
To get the background quiet Sun, we extract the mean ($\overline{C}$) and standard deviation ($\sigma$) from the distribution to select all the pixels within the range of $\overline{C} \pm k \sigma$, assuming that contrast values of the quiet Sun have a Gaussian distribution. 
This allows us to refine the identification of the background quiet Sun by removing pixels belonging to bright regions from further analysis. The mean ($\overline{C}$) and standard deviation ($\sigma$) are then again calculated from the distribution of these selected pixels (red; \autoref{network_detection}(c)). One such distribution of selected pixels within $\overline{C} \pm k\sigma$ for $k=1.55$ is shown in \autoref{network_detection}(c). In \autoref{network_detection}(d), the mean $\overline{C}$ is plotted as a function of $k$, $\forall~k\in [0.65, 3.0]$, where $k$ is incremented in the step of 0.05. The minimum value $\overline{C}$ ($\overline{C}_{\mathrm{min}}$), called minimum mean contrast, and corresponding ($\sigma_{\mathrm{min}}$), the minimum standard deviation is the representative of background quiet sun intensity distribution.

Once we get the $\overline{C}_{\mathrm{min}}$ and $\sigma_{\mathrm{min}}$ we calculate the threshold as, 

\begin{equation}
T=\overline{C}_{\mathrm{min}} + m_{\mathrm{n}}*\sigma_{\mathrm{min}},
\label{eq:threshold}
\end{equation}

\begin{figure*}[htb!]
 \centering
\includegraphics[width=8cm]{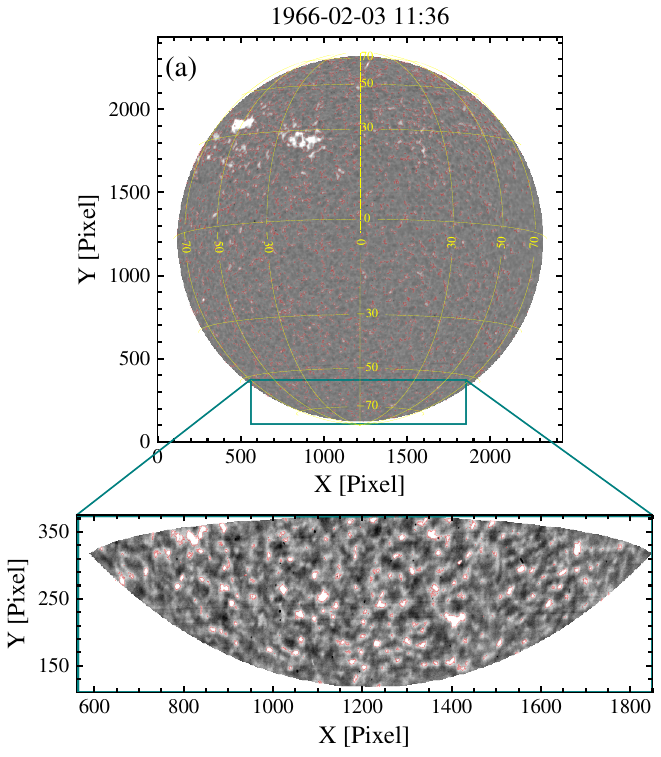}
\includegraphics[width=7.5cm]{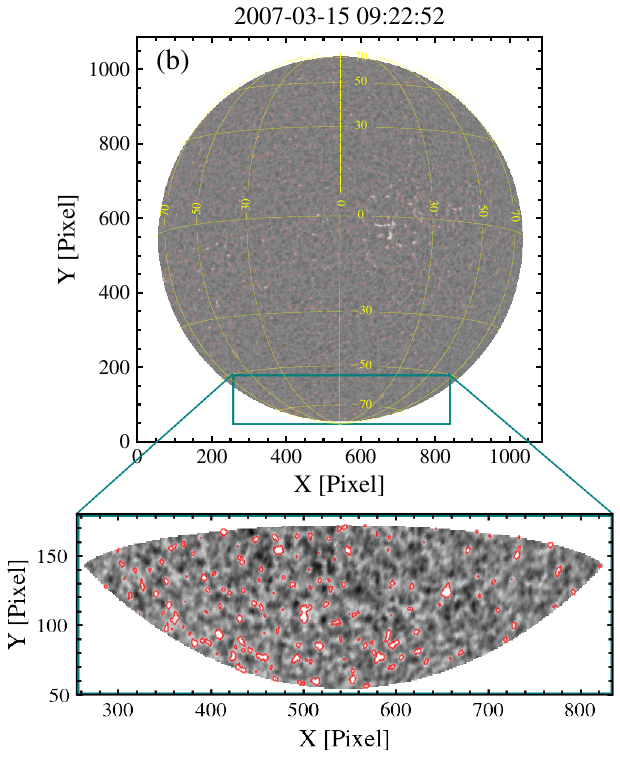}
\caption{(a and b) The red contours depict the detected features, including plages and network regions, within \ca\ image (contrast $\in [-0.5,0.5]$) of KoSO and PSPT-R observed on 1966-02-03 11:36 IST (06:06 UT) and 2007-03-15 09:22:52 UT respectively. This marking follows the application of the threshold derived with Eq. \ref{eq:threshold} (see Sect. \ref{sec:polarnetworkdetection}). The rectangular green box marks approximately the selected region for $-55^{\circ}$ to $-90^{\circ}$ in the south pole (yellow overlaid grid) for the calculation of PNI. The inset (contrast $\in [-0.1,0.1]$) shows the exact region of interest for the PNI estimation covering the south pole of the solar disk, spanning from a latitude of $-55^{\circ}$ to $-90^{\circ}$. Red contours highlight the presence of bright polar network regions within this specified region.}
\label{network_detection1}
\end{figure*}

\noindent{}to identify the network regions in \ca\ observations. Here, $m_{\mathrm{n}}$ is a constant having a value of 3.2 for both  KoSO and PSPT-R \ca\ observations. The value of $m_{\mathrm{n}}$ is first estimated based on the visual inspection of the identified network regions but then set based on the study of the correlation between our PNI estimates and the WSO polar field (see Section \,\ref{reconstruct}). In particular, we considered the value of $m_{\mathrm{n}}$ that returns the maximum value of \cc. 
A representative example of the identified network regions with this NR approach is shown in \autoref{network_detection1}(a) for the KoSO \ca\ observation taken on 1966-02-03 11:36 IST (06:06 UT) and, in \autoref{network_detection1}(b) for PSPT-R \ca\ observation taken on 2007-03-15 09:22:52 UT.



\subsection{Polar Network Index}

To calculate the PNI, we only focus on the network identified close to the northern and southern pole of the Sun within the latitude range of $\pm (55^{\circ}-90^{\circ})$ during August-September (for the northern hemisphere) and February-March (for the southern hemisphere) ensuring the optimal viewing of the northern and southern pole respectively. Such representative examples of the optimal view of the South Pole are shown in the inset box of \autoref{network_detection1}(a and b) for KoSO and PSPT-R, respectively.
The PNI is computed by summing the total number of pixels within these polar network masks within the latitude range of $\pm (55^{\circ}-90^{\circ})$ and expressing them in terms of a millionth of the area of the solar disk, i.e.,

\begin{equation}
    {\rm PNI} = \frac{{\rm Number~of~Network~Pixels}}{\pi R^2}\times10^6,
\end{equation}

\noindent{}where $R$ is the radius of the solar disk in pixels. We normalised the sum of pixels within polar network masks by the area of the solar disk to consider the varying solar disk sizes during the months of consideration, which, to the best of our knowledge, was not considered by \citet{priyal2014} for their PNI estimation. 
We calculate the PNI for all the \ca\ observations from  KoSO (1904\,--\,2007) and PSPT-R (2000\,--\,2022) during the aforementioned months of the year.




\section{Results}\label{Sec:results}


\subsection{PNI Time Series}
\label{sec:polarityassignment}

\begin{figure*}[htb!]
 \centering
\includegraphics[width=\textwidth]{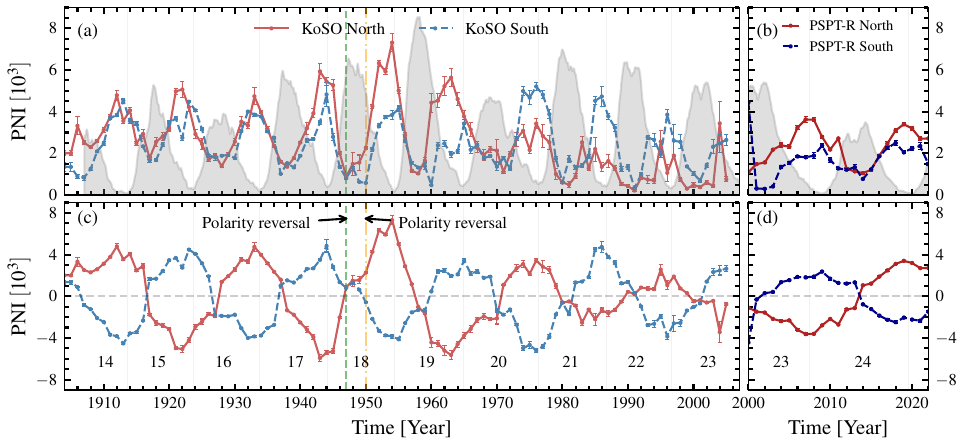}
\caption{Top: The variation of PNI from KoSO (a) and PSPT-R (b) \ca\ data for both northern and southern hemispheres with error bars corresponding to \sse\ errors in the latitude range of $55^{\circ}$ to $90^{\circ}$. The shaded area shows the variation in the total sunspot number (ISSN) scaled to bring into the range of PNI values. Bottom: The PNI series with signed polarity (see the subsection~\ref{sec:polarityassignment} for details on how we assigned the polarity) from KoSO (c) and PSPT-R (d) \ca\ data is shown for both the northern and southern hemispheres. An example of a polarity reversal for the northern and southern hemispheres is shown by dashed green and dash-dot orange lines, respectively, in (a and c).}
\label{pniseries}
\end{figure*}

A two-month averaged PNI time series is constructed for the northern and southern hemispheres by considering the data from August--September and February--March for the respective hemispheres.  In \autoref{pniseries}(a) and \ref{pniseries}(b), we plot the average PNI as a function of the year along with the \sse\ in the mean, for KoSO (1904\,--\,2007) and PSPT-R (2000\,--\,2022) respectively.  In \autoref{pniseries}(a), we note a systematic decrease and larger fluctuations in PNI for KoSO data after 1980, which is due to the degradation of data quality in KoSO in the later half of the 20th century, as identified by \citet{ermolli2009ApJ,theo2023A&A, Mishra_2024}. This effect is even more severe for the northern hemisphere by the end of the 20th century, when we observe a very low value of PNI in this hemisphere. We note that on top of the systematic degradation in image quality after the 1980s, the number of observations is significantly lower in the months of August to September, likely due to the rainy season and monsoon in Kodaikanal. On the other hand, the PNI calculated from PSPT-R shows more consistent variation in the later part of this period. 
Overall, the PNI calculated from PSPT-R observations is at comparable levels, albeit potentially slightly lower, than the PNI calculated from KoSO data. 
This is expected based on the nominal bandpass of the images in PSPT-R being broader than that of KoSO \citep{[see][]theo2019SoPh}; this difference agrees with other results achieved by comparing bright features identified in KoSO and PSPT-R images.
Additionally, we note that the temporal variation of PNI (as with other measures of the polar field strength) exhibits an antiphase relationship with the international smoothed sunspot number v2.0\footnote{Sunspot data from the World Data Center SILSO, Royal Observatory of Belgium, Brussels (\url{https://www.sidc.be/SILSO/datafiles}).} \citep{Clette2015,clette_recalibration_2023,sidc} that can be seen in \autoref{pniseries}(a and b). While the PNI for cycles 14-18 are fairly symmetric in amplitudes, the cycles since then have notable hemispheric asymmetries.


PNI values are inherently positive; hence, to use them as the proxies of the polar field, we assign them with the sign in both hemispheres as follows. For KoSO PNI before 1976 and PSPT-R PNI, we assign and reverse its sign based on the timing of the minimum of the two-month-averaged PNI series in both hemispheres (refer to the green and yellow vertical line in \autoref{pniseries}(a and c)). However, due to the data quality issue for KoSO after 1976, we used the timing of polarity reversal from the WSO polar field measurement to assign the polarity to KoSO PNI after 1976. 
See \autoref{tab:polarity of PNI} for the comparison of the times of polarity reversal for both $\pm (55^{\circ}-90^{\circ})$ and $\pm (70^{\circ}-90^{\circ})$ based on the WSO polar field and based on the minimum of the PNI series.  
In \autoref{pniseries}(c) and \autoref{pniseries}(d), we show the signed PNI series of KoSO (\pnikso\ hereafter) and PSPT-R (\pnipspt\ hereafter), as a function of time. 


\begin{table*}[htb!]
\centering
\caption{Year of polarity reversal for PNI and \Bwso}
\begin{tabular}{l|cc|l|cc|l}
\hline
\hline

Solar Cycle & \multicolumn{2}{c|}{PNI North} & \Bwso\ North & \multicolumn{2}{c|}{PNI South} & \Bwso\ South\\

& $55^{\circ}-90^{\circ}$ & $70^{\circ}-90^{\circ}$ & & $55^{\circ}-90^{\circ}$ & $70^{\circ}-90^{\circ}$  \\               
\hline

14$^*$ &-- & -- & -- & 1907 & 1907 & -- \\
15$^*$ & 1917 & 1917 &-- & 1917 & 1918 & --\\
16$^*$ & 1928 & 1928 &-- & 1927 & 1928 & -- \\
17$^*$ &1938 & 1938 &-- & 1937 & 1937 & -- \\
18$^*$ & 1947 & 1947  &-- & 1950 & 1949 & -- \\
19$^*$ & 1958 & 1958 &-- & 1960 & 1958 & -- \\
20$^*$ & 1971 & 1971 &-- & 1970 & 1970 & -- \\
21$^*$ & 1981 & 1980 &1980 & 1980 & 1981 & 1981 \\ 
22$^*$ & 1991 & 1991 &1990 & 1991 & 1991 & 1992 \\
23$^{**}$ & 2000 & 2000 &2000 & 2002 & 2001 & 2002 \\
24$^{**}$ & 2014 & 2015 &2013 & 2014 & 2014 & 2014 \\

\hline
\end{tabular}
 \tablecomments{\footnotesize{$^*$ KoSO, $^{**}$ PSPT-R}}
\label{tab:polarity of PNI}
\end{table*}

\subsection{Comparison with Polar Faculae Counts}
\label{sec:faculae_pni overlap}

\begin{figure*}[htb!]
 \centering
\includegraphics[width=\textwidth]{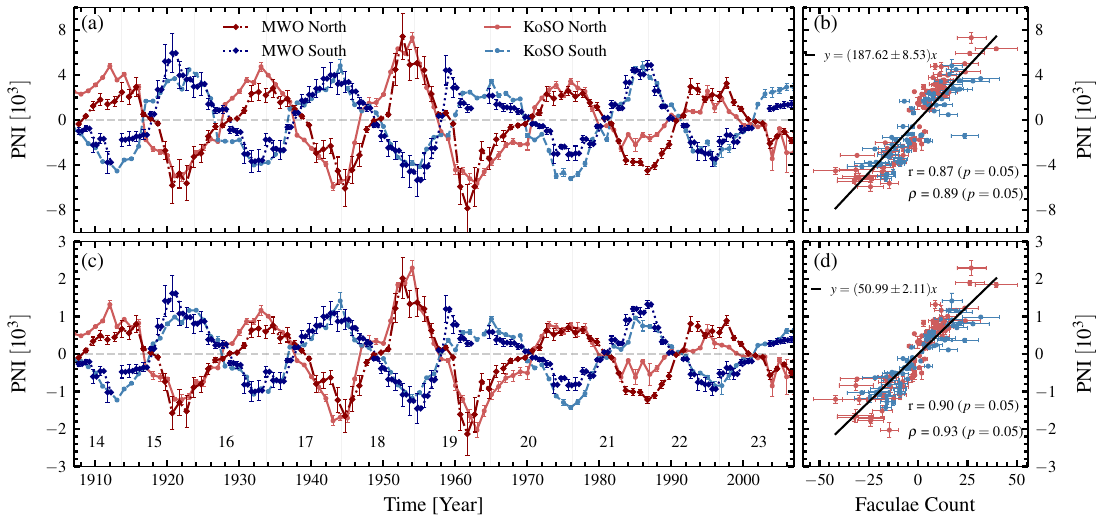}
\caption{Comparison between \pnikso\ with polarity signs in both the northern (filled circle markers solid red line) and southern hemispheres (filled circles markers dashed blue line) to MWO white-light faculae counts in corresponding hemispheres (diamond markers dash-dotted brown line; north and diamond markers dotted navy blue line; south) during the overlapping period from 1907 to 2007 for the latitude range of $\pm (55^{\circ}-90^{\circ})$ (a) and $\pm (70^{\circ}-90^{\circ})$ (c). Scatter plot between \pnikso\ and faculae counts (b and d), which is denoted by red and blue points for the northern and southern hemispheres with corresponding \sse\ error bars. Also shown is a linear fit to the data (black line), noting the fit parameter at the top part of the panel as well as the Pearson ($r$) and Spearman ($\rho$) correlation coefficients.}
\label{pni_faculae}
\end{figure*}

In \autoref{pni_faculae} (a and c), we compare the signed \pnikso\ (1907\,--\,2007) with white-light faculae counts from MWO for latitude range of $\pm (55^{\circ}-90^{\circ})$ and $\pm (70^{\circ}-90^{\circ})$, where the average faculae counts are taken from \citet{munoz2013} and scaled by the factor of 187.62 and 50.99 (calibrated for the period 1907-1980) respectively, to bring them to the scale of \pnikso\ for better comparison. We emphasize here that the faculae count considered in \citet{munoz2013} were for the interval of February 15–March 15 for the south pole and August 15–September 15 for the north pole, whereas in this work, \pnikso\ is considered from August 1 to September 30 for north and February 1 to March 31 for the south. \autoref{pni_faculae} error bars indicate the \sse\ in the mean calculated over the period of consideration for faculae counts and \pnikso. 
Results for the southern hemisphere are, in general, better agreement than those derived from the northern hemisphere except for early observations in 1907--1915  and 1930--1935 covering solar cycles 14 and 16, respectively.
As mentioned above, after 1980, there was a considerable degradation in KoSO \ca\ images, which is very apparent in \pnikso\ as well. Furthermore, in \autoref{pni_faculae}(b) we show a scatter plot between  \pnikso\ and MWO faculae counts for the latitude range from $55^{\circ}- 90^{\circ}$ in the period of 1907\,--\,1980 and also calculate Pearson, $r = 0.87~($significance $p=0.05)$ and Spearman, $\rho = 0.89~(p=0.05)$ correlation coefficient (CC) between them. Similarly, we observe $r = 0.90~($$p=0.05)$ and $\rho = 0.93~(p=0.05)$ for the latitude range of $\pm (70^{\circ}-90^{\circ})$ in \autoref{pni_faculae}(d). The high CC value suggests a strong correlation between them, signifying the importance of PNI as a proxy for the reconstruction of the historical polar field. 
The inclusion of the $\pm (70^{\circ}-90^{\circ})$ latitude range in this specific analysis is due to the fact that polar faculae are predominantly measured within this heliographic latitude range \citep{dyson1923MNRAS,Munoz2012, hovis-afflerbach_two_2022} for the purpose of measuring the polar magnetic field. However, for this work, we primarily use the $\pm (55^{\circ}-90^{\circ})$ latitude range for PNI calculation to align with the WSO polar field measurement latitude range. Consequently, for comparison with polar faculae, we utilize both the $\pm (55^{\circ}-90^{\circ})$ and $\pm (70^{\circ}-90^{\circ})$ latitude ranges to examine the variation in the correlation between PNI and polar faculae.
In addition to \pnikso, similar high values of CCs ($r>0.87$ and $\rho>0.87$) are also found for \pnipspt, and MWO faculae count calculated over the overlapping period of 2000\,--\,2007.


\begin{figure*}[htb!]
 \centering
\includegraphics[width=\textwidth]{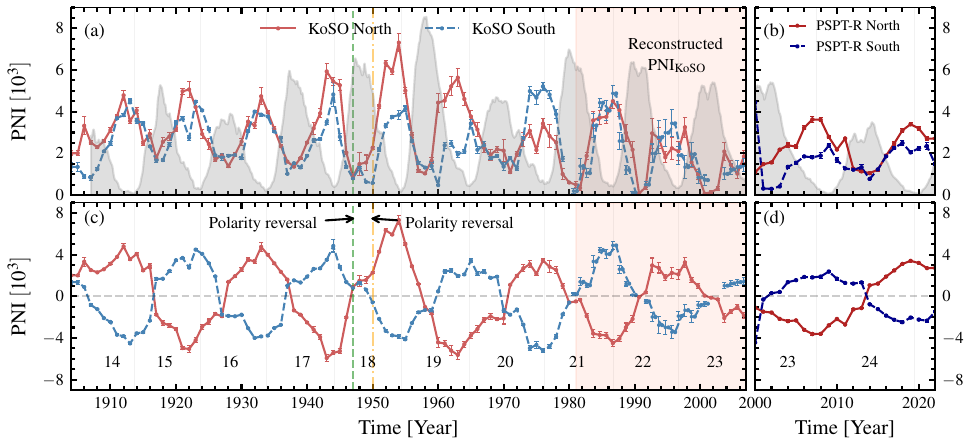}
\caption{Similar to \autoref{pniseries}, except that \pnikso\ has been reconstructed using MWO polar faculae for the period 1981\,--\,2007 (shaded orange region).}
\label{pni_construct}
\end{figure*}

In this section and \autoref{sec:polarityassignment}, we addressed the underestimated counts of \pnikso\ for the northern hemisphere, which were caused by the degradation in data quality of KoSO \ca\ images. To mitigate this issue, we reconstructed \pnikso\ for the period from 1981 to 2007 using a scaling factor of 187.62, as derived above for MWO polar faculae counts. This approach allowed us to produce an enhanced version of the PNI, combining both \pnikso\ and MWO polar faculae, as illustrated in \autoref{pni_construct}. We will use these reconstructed PNI for the analysis in the following sections.



\subsection{Reconstruction of Polar Field}
\label{reconstruct}

Now, we discuss the main goal of the work, which is to reconstruct the polar field ($B_{\rm P}$) over the last century from \pnikso\ and \pnipspt. We start with calculating the Pearson ($r$) and Spearman ($\rho$) correlation coefficients for \pnikso\ and \pnipspt\ with both WSO polar magnetic field (\Bwso) and polar field derived from AFT model (\Baft), which is summarized in \autoref{tab:ccPNIB}. Based on these correlation coefficients, we undertake the reconstruction using a linear model $B_{\rm P}\propto {\rm PNI}$, i.e. $B_{P} =b({\rm PNI})$. We also calculate the coefficient of determination ($R^{2}$-score) for the optimal fitting parameters calculated using the least square fitting, and we find that the $R^{2}$-score for \pnikso\ and \pnipspt\ with WSO polar field (\Bwso) are 0.92 and 0.86, respectively, indicating a strong linear relationship between these two physical quantities. Furthermore, we also test our model for \pnikso\ and \pnipspt\ with \Baft\, which yield the $R^{2}$-score of 0.83 and 0.82, respectively. 

\begin{table}[htbp!]
\caption{Correlation coefficients between PNI and polar field series considered in this study}
\begin{tabular}{l|cc|cc}
\hline
\hline
& \multicolumn{2}{c|}{\Bwso} & \multicolumn{2}{c}{\Baft}\\               
& $r$ &$\rho$ & $r$ &$\rho$\\
\hline
\pnikso & 0.96 & 0.96 & 0.92 & 0.94\\
\pnipspt & 0.93 & 0.89 & 0.91 & 0.85\\
\hline


\end{tabular}
\tablecomments{\footnotesize{\pnikso\ and \pnipspt\ are PNI of KoSO and PSPT-R, respectively, $r$ and $\rho$ are the Pearson and Spearman correlation coefficients, respectively.}}
\label{tab:ccPNIB}
\end{table}



\begin{figure}[htb!]
 \centering
\includegraphics[width=\columnwidth]{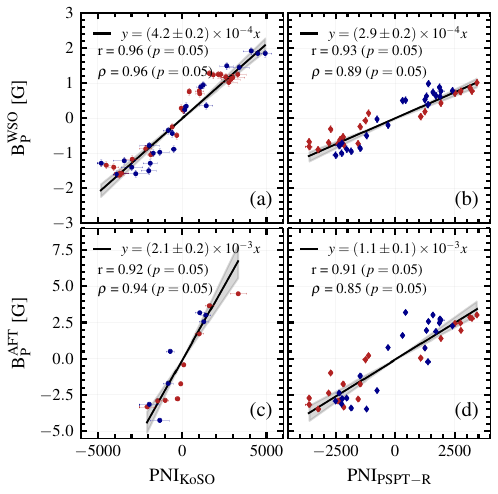}
\caption{Scatter plots between PNI and \Bp. In particular, we show the pairs: (a) \Bwso\ and \pnikso\ for the period of 1977 to 2000, (b) \Bwso\ and \pnipspt for 2001\,--\,2022, (c)\Baft\ and \pnikso\ for the period of 1996 to 2007 and (d) \Baft\ and \pnipspt\ for the period of 2001 to 2022, along with their corresponding error in measured PNI and polar field. Results for the northern hemisphere are shown with red diamond (PSPT-R) and red circle (KoSO) symbols, while those for the southern hemisphere are displayed with blue diamond (PSPT-R) and blue circle (KoSO) symbols. The best-fit lines are shown with the solid black line, whereas shaded regions represent the 95\% confidence interval. In case \pnipspt\ ($<2$) and \Baft\ ($<0.02$), error bars are very small, and hence, they barely appear in the plot. Also listed are the Pearson ($r$) and Spearman ($\rho$) correlation coefficients.}
\label{pni_calibration}
\end{figure}

Firstly, we use the \Bwso\ to get $b$, the calibration constant for \pnikso\ and \pnipspt. We restricted ourselves to the period from 1977 to 2000 in the case of \pnikso\ due to the concern about less reliable data beyond this time frame. In \autoref{pni_calibration}(a), we show a scatterplot between \Bwso\ with \pnikso\ and fit our linear model, which yields the calibration constant $(4.2\pm0.2) \times 10^{-4}$ (95\% confidence interval), represented by the solid black line in \autoref{pni_calibration}(a). Similarly, in \autoref{pni_calibration}(b), we plot \Bwso\ with \pnipspt\, for the period of 2001\,--\,2022, which gives the $(2.9\pm0.2) \times 10^{-4}$. We limit the analysis to the period 2001\,--\,2022 because the PSPT-R data are more homogeneous over this period \citep{ermolli10.3389}.
Following that, we have used \Baft\ to calculate the $b$ for both \pnikso\ (1996\,--\,2007) and \pnipspt\ (2001\,--\,2022), which results in $(2.1\pm0.2) \times 10^{-3}$ and $(1.1\pm0.1) \times 10^{-3}$ respectively. We also observe a significant correlation ($r>0.91$ and $\rho>0.85$) between PNI and \Baft\ for both PSPT-R and KoSO data.

\begin{figure*}[htb!]
 \centering
\includegraphics[width=\textwidth]{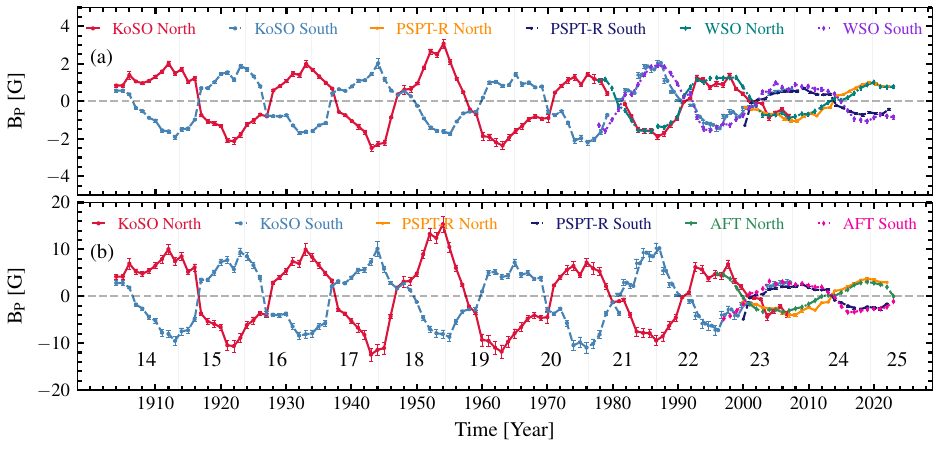}
\caption{(a) The polar magnetic fields obtained from \pnikso\ (solid red and dashed blue for the northern and southern hemisphere, respectively) and \pnipspt\ (solid orange and dashed navy for the northern and southern hemisphere, respectively) spanning the time interval from 1904 to 2022 are shown, alongside \Bwso\ for the overlapping timeframe (1976\,--\,2022) in the northern (dash-dot dark green) and southern (dotted purple) hemispheres. (b) The polar fields obtained from \pnikso\ and \pnipspt\ using calibration value (with \Baft; 1996\,--\,2007 and 2001\,--\,2022 respectively) are compared with \Baft\ (1996\,--\,2023) of the northern (dash-dotted green) and southern (dotted pink) hemispheres, respectively.}
\label{polar_field_koso}
\end{figure*}

We reconstruct polar fields using \pnikso\ (reconstructed for 1981\,--\,2007) and \pnipspt\ based on the calculated value of $b$ for each case of \Bp\ (WSO and AFT) individually for the northern and southern hemispheres. In \autoref{polar_field_koso}(a) and \ref{polar_field_koso}(b), we plot the constructed \Bp\ as a function of time-based on \Bwso\ and \Baft\ respectively for their complete period of observations. Furthermore, the polar fields derived from PSPT-R for both hemispheres are in excellent agreement with \Bwso\, underscoring the robustness of this methodology in deriving polar fields from the two \ca\ PNI series. 
During the overlapping period, the polar field estimated from KoSO and \Bwso\ also shows a good agreement for both hemispheres.
The error bars for the polar fields, shown in \autoref{polar_field_koso}(a), are derived by applying standard formulae for the propagation of relative uncertainties 
$\frac{\sigma_{y}^2}{y^2}=\frac{\sigma_{b}^2}{b^2}+\frac{\sigma_{x}^2}{x^2}$ where $x$ and $y$ are the equivalent to \Bp\ and PNI respectively, and $\sigma$ represents the corresponding uncertainty in them, at the same time $b$ and $\sigma_b$ are the calibration constant and fitting error corresponding to it.

\subsection{Composite PNI Series}

\begin{figure*}[htb!]
 \centering
\includegraphics[width=\textwidth]{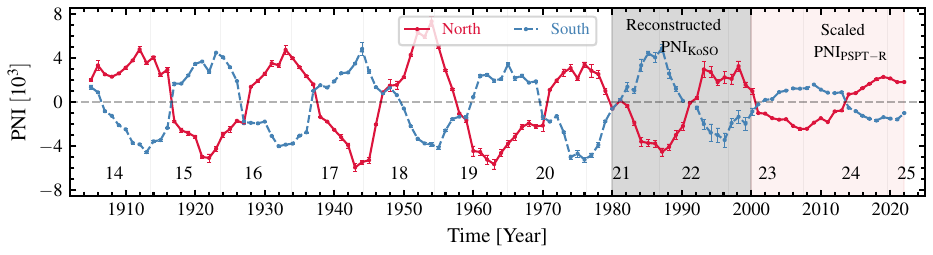}
\caption{The composite PNI series (1904\,--\,2022) merges results from \ca\ data from KoSO and PSPT-R. The grey and light pink shaded regions represent the reconstructed \pnikso, derived from faculae counts from MWO and scaled \pnipspt\ using \pnikso\ for the periods 1980\,--\,2000 and 2001\,--\,2022, respectively.}
\label{pni_composite}
\end{figure*}

We construct a composite PNI series by integrating \pnikso\ and \pnipspt\ data. Specifically, we include all \pnipspt\ data from 2000 onwards and all \pnikso\ data prior to 1980. 
Our composite PNI series, displayed in \autoref{pni_composite}, is further refined to achieve uniformity across the entire 119-year period by calibrating \pnikso\ and \pnipspt\ data over the overlapping period from 2000 to 2007. Due to the unreliability of KoSO data for the interval of 1980 to 2000, we use rescaled polar faculae counts from MWO data to reconstruct \pnikso\ (1980\,--\,2000, see the grey shaded region in \autoref{pni_composite}), adjusting by the 187.62 factor derived from the 1904\,--\,1980 data comparison in \autoref{sec:faculae_pni overlap}. From this reconstructed \pnikso, we perform a linear fit ($y = bx$) to \pnipspt\ data for the same period, determining a calibration factor of 0.67. This factor is then used to scale \pnipspt\ data to align with \pnikso, resulting in the rescaled \pnipspt\ values shown in the pink shaded region of \autoref{pni_composite}. Consequently, we establish a long-term PNI composite series spanning 1904 to 2022, which will be made public to the community.

\subsection{Polar Field and Amplitude of Solar Cycle}

The sunspot number series introduced by R. Wolf in 1849 \citep{wolf1856} and currently compiled by WDC-SILSO, as the international sunspot number \citep[ISNv2;][]{clette_recalibration_2023}, is the most widely recognized proxy of the solar cycle. The strength and timing of the solar cycle are typically defined based on these numbers. It has been noted that the polar field at the time of solar minima is one of the best proxies for predicting the following solar cycle \citep{upton2023JGRA}. Therefore, we used the newly constructed polar field data over the last 11 solar cycles (Cycle~14 to Cycle~24) to estimate the strength of Solar Cycle~25, which seems to have entered into the maximum phase in 2024 \citep{jha2024}. As we are very close to the maximum of Cycle~25, such a prediction serves mostly as a performance test of our newly reconstructed series of polar magnetic fields. 

In this study, we utilize the 13-month smoothed sunspot number v2.0 (ISSIv2.0) to determine the timing of cycle minima and the amplitude ($A_{\rm Max}$) of each solar cycle. Additionally, we use \Bp\ data obtained from both \pnikso\ and \pnipspt\ to estimate the polar field at solar minimum ($B_{\rm P}^{\rm Min}$) for ten solar cycles (15–24). Here, $B_{\rm P}^{\rm Min}$ refers to the combined average of polar field values for the northern and southern hemispheres at the minimum preceding the $A_{\rm Max}$ of a specific solar cycle.

\begin{figure}[htb!]
 \centering
\includegraphics[width=6cm]{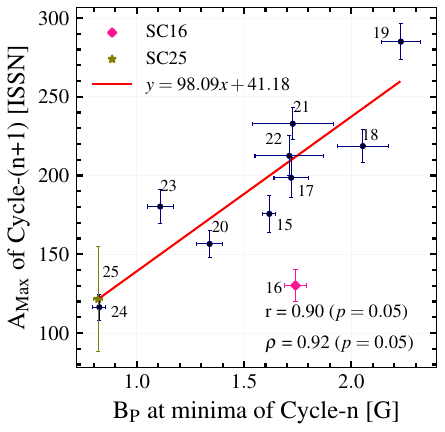}
\caption{ The correlation analysis illustrates the relationship between $A_{\rm Max}$ of Cycle-(n+1) [ISSN] and the $B_{\rm P}^{\rm Min}$ at the minima of Cycle-n, represented in navy blue colour excluding the solar cycle 16 (pink point). A linear regression model ($y=bx+c$), denoted by the red line, has been applied. The estimated strength of solar cycle 25 is presented in dark green point.}
\label{cycle_corr}
\end{figure}

We conduct a correlation analysis between $B_{\rm P}^{\rm Min}$ and $A_{\rm Max}$, yielding a correlation coefficient of $r = 0.90$~ ($p = 0.05$) and $\rho = 0.92$~($p = 0.05$). This analysis excludes Cycle 16 based on the best possible correlation, marked (pink) in \autoref{cycle_corr}. This observation was also seen in past works utilizing polar networks \citep{priyal2014} and polar faculae \citep{rodr2024SoPh}, but the reason for Solar Cycle 16 being an outlier still remains unknown.
Some of the potential reasons for this can be issues with the sunspot number series or the \ca data employed here.
We note that a recent recalibration of the sunspot number series over cycle 16 by \citet{bhattacharya_rudolf_2024} did not reveal any potential issues with the sunspot number series over that period.
Further work is required to understand this issue.


A linear fit ($y=bx+c$) is performed in \autoref{cycle_corr}, yielding a slope and intercept of $b = 98.1 \pm 18.2$ and $c = 41.2 \pm 30.0$, respectively. Using this calibration value and the polar field estimated from \pnipspt\ at the previous solar cycle minimum, we estimate the strength of solar cycle 25 to be $121.6 \pm 33.1$, which is also shown in 
\autoref{cycle_corr} with dark green marking.
Our estimate for the amplitude of solar cycle 25 is slightly lower than the current trends (about 130 - 140) and previous predictions \citep{bisoi2020JGRA,upton2023JGRA,JAVARAIAH20241518}.
However, within the uncertainty of our estimate, it is consistent with the others. 




\section{Summary and Conclusion}\label{Sec:summary}

We utilized century-long \ca\ data from KoSO (1904\,--\,2007) and  \ca\ data from PSPT-R (2000\,--\,2022) to investigate the polar network regions. From the detected polar network, we calculated PNI within the latitude range of $\pm (55^{\circ}-90^{\circ})$ for both the northern and southern hemispheres, thereby constructing PNI series for the period 1904\,--\,2022 from both these data set. We get polarity-signed PNI values based on the minimum PNI time series averaged in two months. We found that the \ca\ PNI show a strong correlation with MWO polar faculae (\autoref{pni_faculae}), validating the potential of the polar network as a proxy of the polar field. In addition, we utilized these polar faculae counts to fill the data gap for \pnikso due to their degrading quality after 1980. Also, we combined KoSO \ca\ data with PSPT-R \ca\ observations to extend the PNI series to cover the period from 1904 to 2022. 

The composite PNI series, derived from combining KoSO and PSPT-R observations, enabled us to reconstruct the historical polar magnetic field from 1904 to 2022. The reconstructed polar field based on \pnipspt\ showed excellent agreement with both the WSO and AFT polar fields (see \autoref{polar_field_koso}), underscoring the efficacy of using the polar network as a proxy for polar field reconstruction. The reconstructed PNI series data for the entire period are available to the community through an open and public repository (GitHub+Zenodo), which can be accessed at \url{https://doi.org/10.5281/zenodo.14676548} \citep{dibya_kirti_2025_14676549}.
Furthermore, we used the polar field derived from PNI to estimate the relation between the amplitude of the solar cycle and the strength of the polar field at the preceding cycle minima. Furthermore, we used this relation to compare the amplitude of the current cycle with other predictions.
This reconstructed \Bp\ over the last $\approx$11 solar cycle from historical data is a valuable asset for constraining the solar dynamo and SFT models, which were limited due to the unavailability of historical polar field data. To further complete the PNI series, particularly for the period from 1980 to 2000, where data quality is a concern, future efforts should focus on integrating additional \ca\ data sources to construct a more comprehensive composite PNI series.

\begin{acknowledgments}
We express our gratitude to the observers at the Kodaikanal Solar Observatory and the individuals involved in the digitization process for their efforts in providing extensive solar data spanning over a century to the scientific community. Kodaikanal Solar Observatory is a facility of the Indian Institute of Astrophysics, Bangalore, India. \ca\ raw data are now available for public use at \url{http://kso.iiap.res.in} through a service developed at IUCAA under the Data Driven Initiatives project funded by the National Knowledge Network. We acknowledge WDC-SILSO, Royal Observatory of Belgium, Brussels, for the sunspot data. Additionally, we extend our sincere gratitude to Rome/PSPT \url{https://www.oa-roma.inaf.it/pspt-daily-images-archive/} for providing the easily accessible data that we have utilized in our current work. The funding support for DKM's research is from the Council of Scientific \& Industrial Research (CSIR), India, under file no.09/0948(11923)/2022-EMR-I. T.C. acknowledges funding from the European Research Council (ERC) under the European Union's Horizon 2020 research and innovation programme (grant agreement No. 101097844 — project WINSUN). This study has made use of SAO/NASA Astrophysics Data System's bibliographic services.

\end{acknowledgments}

\section{Data availability statement}
PNI composite series combining KoSO and Rome/PSPT Ca II K data and polar field derived from these are available at GitHub and Zenodo repository through \url{https://doi.org/10.5281/zenodo.14676548} \citep{dibya_kirti_2025_14676549}.

\bibliography{references}{}
\bibliographystyle{aasjournal}

\end{document}